\documentclass{an}
\usepackage{graphicx}
\usepackage{times}
\Volume{0}               
\Year{2004}              
\Pagespan{000}{000}      

\begin{document}
\lhead[\thepage]{C.A. Watson \& V.S. Dhillon: Roche tomography of the secondary stars in CVs}
\rhead[Astron. Nachr./AN~{\bf 325}, No. \volume\ (\yearofpublic)]{\thepage}
\headnote{Astron. Nachr./AN {\bf 325}, No. \volume, \pages\ (\yearofpublic) /
{\bf DOI} 10.1002/asna.\yearofpublic1XXXX}

\title{Roche tomography of the secondary stars in CVs}

\author{C.A. Watson and V.S. Dhillon}
\institute{Department of Physics
and Astronomy, University of Sheffield,  Sheffield S3 7RH, UK}
\date{Received {date}; 
accepted {date};
published online {date}} 

\abstract{The secondary stars in cataclysmic variables (CVs) are key
to our understanding of the origin, evolution and behaviour of this
class of interacting binary. In seeking a fuller understanding of
these objects, the challenge for observers is to obtain images of the
secondary star. This goal can be achieved through Roche tomography, an
indirect imaging technique that can be used to map the
Roche-lobe-filling secondary. The review begins with a description of
the basic principles that underpin Roche tomography, including methods
for determining the system parameters. Finally, we conclude with a
look at the main scientific highlights to date, including the first
unambiguous detection of starspots on AE Aqr B, and consider the
future prospects of this technique.  \keywords{binaries: close --
novae, cataclysmic variables -- stars: late-type -- stars: imaging --
stars: spots} } \correspondence{c.watson@sheffield.ac.uk}

\maketitle

\section{Introduction}

Cataclysmic variables (CVs) are defined as semi-detached binary
systems consisting of a Roche-lobe-filling secondary star (typically a
lower main-sequence star) transferring material to a white
dwarf primary star. The term {\em cataclysmic} refers to
their violent but non-destructive outbursts, which occur with a wide
range of frequencies and amplitudes. Non-magnetic CVs accrete material
onto the primary star via a gas stream and accretion disc, and at the
position where the gas stream impacts upon the disc a shock-heated
region of enhanced emission, called the bright spot, is formed.

In the presence of a strong magnetic field from the white dwarf, the
nature of the accretion process changes dramatically. Under these
conditions the white dwarf will accrete material along the field lines
either directly from the gas stream or via a truncated disc. These
latter systems are called {\em polars} or {\em intermediate polars},
respectively, and collectively they form the magnetic CVs. Typically,
the separation of the stellar components in CVs is $\sim$ 1R$_{\odot}$
and their orbital periods are of the order of a few hours.

\subsection{Why image the secondary stars in CVs?}
\label{sec:why}

Surface images of the secondary stars would have far-reaching
implications for CV studies. For instance, conventional methods of
measuring the binary parameters using the secondary star absorption
lines assume that the centre-of-mass and centre-of-light of the
secondary coincide. This is often not the case as, for many of these
systems, the secondaries are known to be irradiated and hence such
measurements will be prone to systematic errors. With the availability
of surface images any irradiation pattern on the secondary would be
known, allowing an accurate determination of the binary
parameters. The study of the irradiation pattern itself may also yield
clues to the geometry of the accretion structures around the white
dwarf - such as the vertical extent of the accretion disc.

Perhaps most importantly of all, images could be used to identify
features related to magnetic activity on the secondary, such as cool
starspot regions. CV secondaries are rapidly rotating and, in the
presence of a dynamo action, are expected to be magnetically
active. Indeed, magnetic activity cycles on the secondary star have
been invoked to explain variations in their orbital periods, mean
brightness and mean outburst intervals. Furthermore, the magnetic
field of the secondary is thought to play a vital role in the
evolution of the binary as a whole, allowing angular momentum to drain
from the binary along the field lines via the magnetic braking
mechanism. This drives the system to shorter orbital periods and
sustains mass transfer between the components. Images would,
therefore, lend evidence for the existence of magnetic activity and,
in turn, the magnetic braking theory. In addition, the number, size,
variability and distribution of starspots as deduced from surface
images would provide key tests of stellar dynamo theories under
extreme environmental conditions.

\section{The principles of Roche tomography}
Unfortunately, the secondary stars in CVs have typical radii of 400
000 km and are located at distances of 200 parsecs. This means that to
detect a feature covering half of the secondary would require a 10
000-m class telescope operating at the diffraction limited resolution
- well beyond what is technically achievable for now. Since direct
imaging has been rendered impossible for the foreseeable future,
astro-tomographic techniques must be employed in order to achieve the
micro-arcsecond resolution required for imaging purposes. Roche
tomography is one such technique and in this section we shall describe
its basic principles and how it can be used to determine the binary
parameters.

\begin{figure}
\begin{center}
\resizebox{7.7cm}{!}
{\includegraphics[angle=-90.]{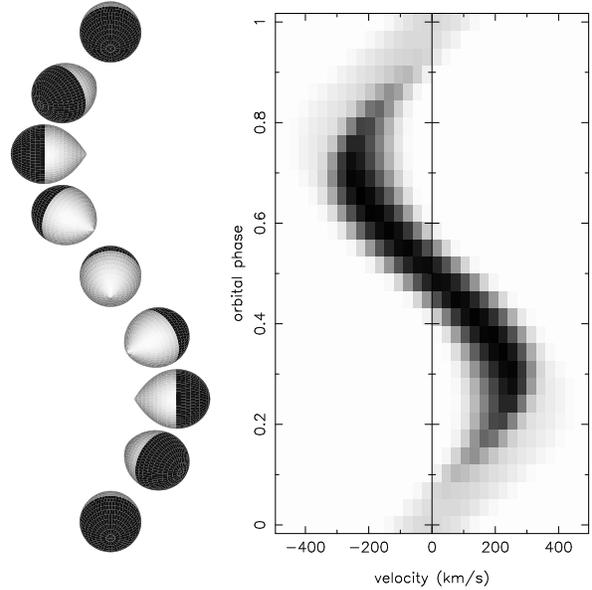}}
\caption{The principles of Roche tomography. By studying the profiles one can
ascertain information on the secondary star's centre-of-mass (from the
movement of the profiles), projected radius (from the variations in $v\sin i$)
and surface features (from intensity variations).}
\label{fig:principles}
\end{center}
\end{figure}

In Roche tomography (Rutten \& Dhillon, 1996) the secondary star is
modelled as a grid of quadrilateral tiles or surface elements with
approximately equal areas, all lying on the critical potential surface
defining the Roche lobe.  Each surface element is then assigned a copy
of the local specific intensity profile convolved with the
instrumental resolution. These profiles are scaled to take into
account the intensity of each element, and also phase dependent
effects such as variations in the projected area, obscuration, and
limb-darkening. The contribution from each element is then
Doppler-shifted according to its radial velocity at any particular
phase (assuming that the secondary star is synchronously
rotating). Finally, the rotationally broadened profile of the
secondary star at that phase can be simply calculated by summing up
the contributions from the surface elements over the visible
hemisphere. An example of this forward process is shown in
Fig.~\ref{fig:principles}.

When reconstructing an image of the secondary star the reverse process
from what has been described above is carried out. In this case, the
strengths of the profiles contributed from each tile are iteratively
adjusted until an image is obtained whose predicted data is consistent
with the observed data (i.e. $\chi^2$=1)\footnote{Note: minimising
$\chi^2$ does not give a good solution since this results in an image
dominated by noise.}. Since there are an infinite number of images
that can fit the data equally as well, it is necessary to employ a
regularisation statistic in order to select a unique image. Following
Horne (1985), we select the image with maximum entropy relative to
some default map.

Essentially, the default map contains prior information about the
intensity distribution across the secondary (if known) and is the
image that the reconstruction will default to in the absence of any
data. In its most non-committal or least informative form each element
in the default map can be set to the average value in the
reconstructed map (a moving-average default) - and the global maximum
of the entropy occurs when the reconstructed and default maps are
identical. If the data are good then the choice of default map makes
little difference to the final image. An efficient algorithm for
maximising entropy subject to the constraint imposed by $\chi^2$ is
given by Skilling and Bryan (1984) and has been implemented by them in
the {\sc fortran} package {\sc memsys}. By application to Roche
tomography a solution is selected which leaves the largest remaining
uncertainty (i.e. the maximum entropy) which is consistent with the
data constraints. In this way, the fewest additional assumptions or
biases are introduced. 

\subsection{Techniques}

As discussed in Section~\ref{sec:why}, if the centre-of-mass and
centre-of-light of the secondary are not coincident then the star's
observed radial-velocity curve will be distorted from the pure sine
wave that represents the true motion of its centre-of-mass. This
occurs as a result of the geometrical distortion of the secondary star
and the presence of surface inhomogeneities, particularly
irradiation. Conventional methods of measuring the system parameters
do not take these effects into account or, at best, require
implementation of an ad hoc correction (e.g. Wade \& Horne, 1988).

\begin{figure}
\begin{center}
\resizebox{7.0cm}{!}
{\includegraphics[angle=-90.]{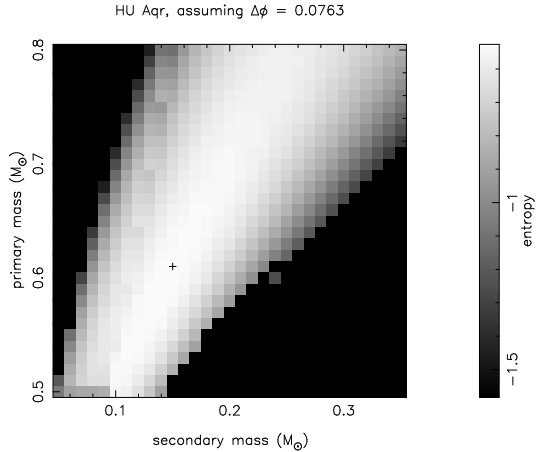}}
\caption{The entropy landscape for HU Aqr. The + marker indicates the
binary parameters that result in the map of highest entropy.}
\label{fig:eland}
\end{center}
\end{figure}

Since Roche tomography automatically takes these factors into
consideration then the system parameters can be obtained which contain
the least systematic bias. In Roche tomography, adopting incorrect
parameters causes spurious artefacts to be introduced during the
reconstruction, normally in the form of equatorial banding (see Watson
\& Dhillon, 2001). This artificially increases the information content
of the image, thereby reducing its entropy. Assuming that the correct
binary parameters are those that minimise the number of artefacts in
the image, they can then be found by searching for the parameters that
result in the reconstruction with the highest entropy. This is shown
graphically as an {\em entropy landscape} (Fig.~\ref{fig:eland}) where
the greyscale of each square represents the entropy value obtained for
the respective masses during image reconstruction.  In similar
fashion, this method can be extended in order to obtain the systemic
velocity and inclination. Simulations by Rutten \& Dhillon (1994) have
shown that the masses determined by this technique can be accurate to
better than $\sim2$ per cent in the presence of irradiation.

\section{Science highlights}
To date, Roche tomograms of the secondary star in 5 different CVs
exist in refereed publications (see Table~\ref{tab:pubs} for a
list). The first practical application of Roche tomography was to the
nova-like DW UMa observed during a low-state when the secondary was
clearly visible.  By co-adding the narrow components of the H$\alpha$,
H$\gamma$ and H$\delta$ emission lines Rutten \& Dhillon (1994) were
able to show that they originated on the irradiated face of the
secondary, though they were unable to determine the system parameters
in this test case.

\begin{table}
\caption[]{List of all refereed publications containing Roche
tomograms of CVs as of September 30th 2003.}
\label{tab:pubs}
\begin{tabular}{llllll}
\hline
Object & Type & Line(s) & Features & Remarks & Ref. \\ \hline
DW UMa & NL & H$\alpha$,$\gamma$,$\delta$ & f & Test case & 1\\
AM Her & P  & Na{\sc i} & t & Test case & 2\\
     &  & Na{\sc i} & t & Full analysis & 3\\
IP Peg & DN & Na{\sc i} & f/l & Full analysis & 3\\
HU Aqr & P  & He{\sc ii} & t & Full analysis & 3\\
QQ Vul & P  & Na{\sc i} & f/t & Full analysis & 3\\
\hline \\
\end{tabular}\\
\begin{flushleft}
\vspace{-0.6cm}
Type codes: DN = dwarf nova; NL = nova-like; P = polar.

Feature codes: (f) irradiation of front (L$_1$ point) of secondary,
(l) irradiation of leading hemisphere, (t) irradiation of trailing
hemisphere, (f/l or f/t) irradiation of L$_1$ point, but stronger
towards leading or trailing hemisphere.

Ref. codes: 1 = Dhillon \& Rutten (1994); 2 = Smith \& Davey (1996);
3 = Watson et al. (2003a)
\end{flushleft}
\end{table}

Derivation of the masses using Roche tomography was first accomplished
by Rutten \& Dhillon (1996) who constructed an entropy landscape
(albeit for a fixed inclination and systemic velocity) for the dwarf
nova IP Peg. The entropy landscape method was then extended to include
a search for the systemic velocity and inclination by Watson et
al. (2003a). This study showed that systemic velocity
measurements derived from fitting radial velocity curves are
unreliable, and that the systemic velocity can be more accurately
determined using Roche tomography.

Furthermore, although it was not possible to find a unique inclination
and, therefore, reliably constrain the component masses of the
non-eclipsing systems in the study\footnote{Note: acceptable changes
in the binary parameters did not greatly affect the derived images of
the secondaries in this study.}, consistent results for the mass-ratios
were found independently of the inclination. A similar analysis of two
separate observations of the X--ray binary Her X--1 (Watson et
al. 2003b) was, however, able to determine a unique inclination and
set of masses in both instances. The masses obtained for the two
datasets agree to within 2 per cent, providing confidence that the
method delivers accurate and self-consistent results.

\begin{figure}
\begin{center}
\resizebox{5.0cm}{!}
{\includegraphics[]{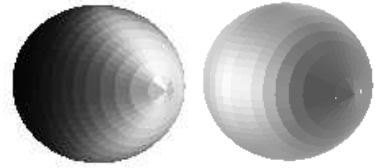}}
\caption{Images of HU Aqr (left) taken in the light of the He{\sc ii}
emission line and IP Peg (right) using the Na{\sc i} doublet, as viewed
at phase 0.6. A moving-average default map was used for both objects.}
\label{fig:twocvs}
\end{center}
\end{figure}

In addition, the resulting images also provide a unique glimpse of the
extreme conditions that exist on the secondary stars
surface. Fig.~\ref{fig:twocvs} shows reconstructed images of the
secondaries in HU Aqr and IP Peg.  The first image presented is that
of the polar HU Aqr in the light of the He{\sc ii} emission line
(Watson et al. 2003a).  The most prominent features are that the
emission (depicted using bright greyscales) originates from the inner
hemisphere and that there is a significant lack of emission from the
leading hemisphere. This can be attributed to irradiation from the
accretion regions which is shielded from the leading hemisphere by the
gas stream and accretion curtain. Likewise, similar irradiation
patterns are also evident in Roche tomograms of two more polars, AM
Her and QQ Vul (see Watson et al. 2003a).

The opposite is apparent in IP Peg (imaged using the Na{\sc I}
doublet, where stronger Na absorption is depicted by bright
greyscales). Here, the strongest impact of irradiation occurs on the
leading hemisphere where the Na{\sc i} is ionised. Since IP Peg is a
non-magnetic CV and therefore harbours an accretion disc, it is
possible that the irradiating source is the bright-spot. This is
located to the correct side of the binary system to illuminate the
leading hemisphere of the secondary.

The tomograms in Fig.~\ref{fig:twocvs} represent the best that can be
achieved using a single absorption or emission line on 4-m class
telescopes. Although this is sufficient to constrain the binary
parameters and map large scale features such as irradiation patterns,
it is insufficient to image small scale features such as starspots. To
do this requires the implementation of techniques such as {\em least
squares deconvolution} (LSD - see Donati et al. 1997) which
effectively stacks the 1000's of stellar lines that can be observed
using an echelle spectrograph to form a single line of high
signal-to-noise.

Here we present the preliminary results of applying LSD and Roche
tomography to the novalike AE Aqr. Observations of AE Aqr were
undertaken on 2001 August 9 \& 10 using the 4.2-m William Herschel
Telescope and the Utrecht Echelle Spectrograph, with simultaneous
photometry taken using the Jacobus Kapteyn Telescope in order to
monitor the flaring behaviour and slit-loss correct the spectroscopic
data.

\begin{figure}
\begin{center}
\resizebox{6.3cm}{!}
{\includegraphics[]{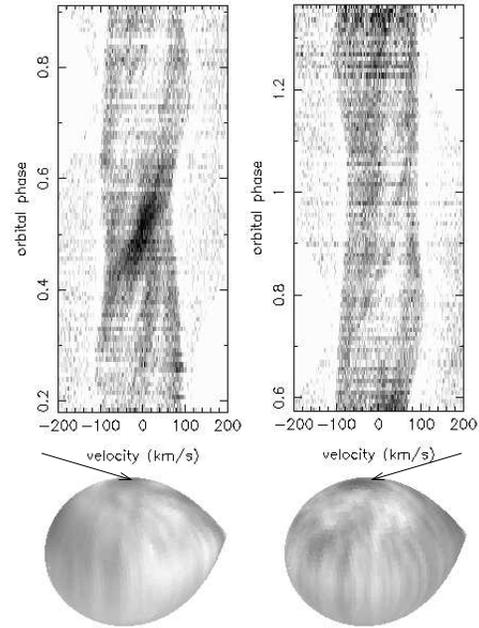}}
\caption{Top panel: Trailed spectra of the LSD profiles of AE Aqr
after removal of the orbital motion and subtraction of a theoretical
profile.  Bottom panel: Preliminary maps of AE Aqr, as the observer
would view it at an inclination of 58$\degr$, constructed from two
consecutive observations. Dark features due to starspots are clearly
visible, especially the spot feature at $\sim$60$\degr$ (arrowed).}
\label{fig:trail}
\end{center}
\end{figure}

The preliminary findings are shown in the top panel of
Fig.~\ref{fig:trail} as trails of the LSD profiles after removal of
the orbital motion and subtraction of a theoretical profile. Clearly
apparent are the expected variations in $v\sin i$ with orbital phase,
as well as the appearance of emission 'bumps' traversing the trails
from blue to red. These 'bumps' are due to the classical distortion of
the line profile by starspots and therefore represent the first
unambiguous detection of starspots on a CV secondary. Further
confidence in the level of the detection can be gained by noting that
several starspot features can be seen to repeat over the two nights.

Roche tomograms for both nights are displayed in the lower panel of
Fig.~\ref{fig:trail}.  Although further work is still required to
sharpen the images, and the rear of the star in night one has not been
observed leading to additional blurring - it is still evident that
several features can be identified in both tomograms. These include
the presence of a high latitude spot extending from $\sim$60$\degr$
towards the polar regions (though not truly polar), and an apparent
band of spots at a latitude close to $\sim$30$\degr$. This is similar
to images of other rapidly rotating stars, where high latitude
starspots have also been found.

\section{Conclusions and future work}

Comparison of the number of Roche tomograms currently in the
literature (Table~\ref{tab:pubs}) with the number of Doppler maps of
CVs shows that Roche tomography is clearly still a fledgling tool. As
such, there are still extensions to the technique that can be
made. One would be to combine Roche and Modulated Doppler tomography
(Steeghs 2003) -- allowing the secondary star and accretion regions to
be mapped simultaneously without many of the constraining axioms of
conventional Doppler tomography.

In addition, as higher quality datasets for CVs become available,
there will be an increasing need to discard the entropy landscape
method. This would be in favour of techniques which allow the binary
parameters to be determined simultaneously during the image
reconstruction process, thereby greatly reducing the computational
burden of such studies. With the current quality of CV datasets,
however, unique and reliable binary parameters may not always exist,
and such methods may produce erroneous results.

Even without these enhanced features, Roche tomography has shown
itself to be capable of constraining the system parameters and
providing an insight into the effects of irradiation in such
objects. In addition, the Roche tomograms of AE Aqr provide the first
conclusive proof that starspots are present on the surface of CV
secondaries.

We can now look forward to imaging CV secondaries in unprecedented
detail, and future work should aim to address questions as to what
role the secondary star magnetic field plays in both the short term
evolution of CVs, such as the effect on accretion rates and cyclical
orbital period variations, as well as the longer term evolution
through magnetic braking. Starspot tracking could also be used to
measure the amount of differential rotation in these stars (e.g.
Cameron \& Donati 1997).  In addition to answering these questions,
such studies will also provide further tests of stellar dynamo
theories under the most extreme environmental conditions.

\acknowledgements The authors thank the efforts of the organising
committee of JD09 of the IAU General Assembly, 2003. Thanks also go to
Tariq Shahbaz for much advice on the AE Aqr data reduction.

\end{document}